\documentclass[review]{elsarticle}

\usepackage{lineno,hyperref}
\modulolinenumbers[5]

\usepackage{amsmath}
\usepackage{graphicx}
\usepackage{caption}
\usepackage{subcaption}
\usepackage{float}
\usepackage{booktabs}
\usepackage{placeins}
\usepackage{multirow}

\journal{Results in Engineering}
\DeclareUnicodeCharacter{2212}{-}
\begin{document}

\begin{frontmatter}

\title{Feature-Level Robustness of Physics-Guided Micro-Doppler Descriptors for classification of Drones and Birds}

\author{Shaiq e Mustafa}
\cortext[cor1]{Corresponding author}
\author{Salman Liaquat\corref{cor1}}
\ead{salman.liaquat@aerospace.pk}
\author{Imran Hafeez Abbasi}
\author{Azhar Hasan}







\begin{abstract}
While micro-Doppler signatures are a proven modality for discriminating between drones and birds, their reliability remains questionable in low-SNR, data-constrained environments where deep learning models fail. This paper presents a systematic analysis of ten statistical and physics-motivated handcrafted features for micro-Doppler classification under controlled signal degradation, using a publicly available 77~GHz frequency modulated continuous wave (FMCW) radar dataset. Micro-Doppler spectrograms are corrupted using Additive White Gaussian Noise (AWGN), phase noise, and their combined effects across a range of $-10 \text{ dB}$ to $10 \text{ dB}$ signal-to-noise ratios (SNRs) and 1\textdegree - 10\textdegree { }phase noise levels. Features extracted from the raw and corrupted spectrograms are evaluated via performance metrics within a stratified 5-fold cross-validation framework for a Support Vector Machine (SVM) and Random Forest classifier, with fixed hyperparameters across all noise levels chosen through grid search. Feature relevance under noise is understood through the lens of permutation based importance calculated as the decrease in macro-averaged F1 score induced by random shuffling of a single feature while keeping all other features fixed. For noise-free data mean classification accuracies of 0.916 $\pm$ 0.095 for the SVM and 0.916 $\pm$ 0.060 for the Random Forest classifier were achieved with F1 scores of 0.909 and 0.892. Experimental results reveal that entropy-based and side-lobe features maintain stable discriminative performance under severe noise, reporting macro-averaged F1 scores of $0.773$ and $0.831$ for the SVM and Random Forest, respectively. Feature-wise evaluation and permutation-based importance analysis reveals that certain features retain complementary discriminative power even when their standalone importance appears low. These findings highlight the importance of principled feature design and provide insight into feature vulnerability and resilience, providing a benchmark for robust, interpretable feature selection in next-generation radar classification systems.

\end{abstract}


\end{frontmatter}

\section{Introduction} 
In recent years, the use of drones, otherwise known as unmanned aerial vehicles (UAVs), has increased significantly due to their low operational cost and surveillance efficiency. Drones have seen a significant increase in their range of applications, such as environmental monitoring, search and rescue missions, and disaster management, among many others. Along with the benefits, concerns regarding public privacy and safety have increased the importance of real-time drone detection and their distinction from birds \cite{white2023, nucum2025micro}, and it is an important research area gaining traction \cite{soares2025use, park2024classification}. 
Traditional means of detection face difficulties due to target's small radar cross section (RCS), low altitude flight profiles, and erratic motion paths, all of which complicate tracking flight paths \cite{seidaliyeva2023advances}. Detection methods based on radio frequency and acoustic information have shown promising results in detecting targets \cite{shahbazian2023human,li2025acoustic}. However, both methodologies face challenges such as radio frequency based methods are vulnerable to interference and acoustic signals degrade in noisy backgrounds especially in urban areas. As a result, radar based micro-Doppler analysis has recently gained traction as an alternative method \cite{hanif2022micro, dronesounddetection, 10163767, 9189414, 10655128}. 
Micro-Doppler based methodologies usually involve transforming radar returns into Doppler-based representations to extract information regarding target classes \cite{10982114}. These representations capture dynamic, class-specific signatures that enable classification of drones, birds, maritime targets, and humans \cite{zhang2025lightweight, park2024classification, rahman2024machine}. In the context of drone classification, the presence of birds poses a significant challenge since their relative size deceives detection methods \cite{Zitar2023}. Another major challenge in micro-Doppler methods is the presence of noise that causes the target signal to be enveloped and obscured, making it difficult to extract distinct features for classification.
To overcome these challenges, existing literature employs methodologies that can be broadly classified into three categories; deep learning–based classification methods, hand-made feature extraction techniques, and noise analysis and degradation studies.

Deep learning representations have demonstrated success in different fields such as biomedical imaging, facial recognition, human pose estimation and visual recognition \cite{chen2026wipowersys, ozaltin2025early}. Motivated by this success, similar models have also been trained on micro-Doppler spectrogram images, reporting classification accuracies higher than 90\% and in some cases, utilizing large curated datasets, as high as 99\% \cite{zhang2025lightweight, park2024classification}. These studies emphasize a reduction in computational complexity and inference times for real-time deployment. However, investigations also show that micro-Doppler frequencies disappear in noisy backgrounds, leading to unstable predictions under temporal and Doppler shifts and low SNR \cite{czerkawski2022robustness}. 

Despite the strong performance metrics, these methods suffer from interpretability with respect to the physical motion mechanisms. Because the methods use a data-driven approach to extract features, the models may learn spurious or dataset-specific features rather than physically meaningful descriptors. Experimental conditions are often constrained, and studies do not subject the models to controlled noise conditions representative of realistic radar, limiting insight into robustness and generalization. For example, Park et al. \cite{park2024classification} assume hovering flight conditions and maneuvering drone dynamics are not accounted for. Additionally, these methods often utilize large labeled datasets, which are scarce in radar scenarios. As an alternative method, handcrafted features can be utilized by grounding feature design in governing physics that explicitly encode semantically meaningful motion characteristics. Such features are expected to generalize better for different conditions while improving interpretability under signal degradation.

Handcrafted features derived from micro-Doppler spectrograms retain explicit physical interpretability. Theses spectrograms and their extracted features inherently have distinguishable factors for different classes. For instance, drone-based spectrograms manifest helicopter rotor modulation (HERM) lines representing modulation caused by rotating blades, while bird-based spectrograms tend to show wing flaps at about 5-7 Hz \cite{rahman2018radar}. Exploiting these differences, studies have extracted several features such as Peak Doppler frequency, Full Width at Half Maximum (FWHM), micro-Doppler bandwidth, micro-Doppler periodicity and other statistical or entropy-based features in order to classify targets \cite{rahman2024machine}. These studies use models such as Logistic Regression and different forms of Support Vector Machines (SVM) trained on these extracted features, reporting classification accuracies on the order of 85\%. Despite their interpretability and data efficiency, these studies lack in depth analysis of feature robustness under controlled noise conditions. In particular, Higher-order statistical features are expected to become less reliable under low SNR conditions motivating the need for explicit robustness analysis in real-life radar scenarios. The key observation here is that handcrafted features do not depend on large datasets thus they are data-efficient, but existing literature lacks feature-wise noise robustness characterization under varying noise conditions that are inherent to radar systems. 


Radar systems are inherently susceptible to noise, which degrades micro-Doppler spectrogram representations and directly impacts classification performance. Studies reveal CNNs suffer significantly under noise such as temporal shifts and adversarial examples showing that detection probability converges to 25\% under low SNR conditions \cite{czerkawski2022robustness}. To mitigate this effect, studies incorporate denoising techniques or evaluate performance metrics as a function of SNR levels. Denoising is often introduced as a preprocessing step, to improve the robustness of the classifier in low SNR conditions \cite{nguyen2024improving}. While this does improve the classification metrics, the primary focus remains signal quality rather than an explicit characterization of how features respond to degradation. Similarly, handcrafted features like continuous wavelet transform (CWT) and the smooth pseudo Wigner–Ville distribution (SPWVD) when coupled with SVM classifier demonstrate an accuracy of 88.63\% even at -8 dB SNR \cite{electronics12244981}. However, these methods remain model-centric and do not isolate the vulnerability or stability of specific features under controlled noise injection.

While existing literature prioritizes maximizing classification accuracy, the intrinsic reliability and variance under noise of the underlying features remain largely under-explored. This investigation moves beyond model-centric evaluation and explicitly characterizes feature vulnerability as a function of corruption to aid classification tasks. To the best of our knowledge, no systematic permutation-based robustness analysis of physics-motivated micro-Doppler features across multiple radar noise mechanisms has previously been reported for micro Doppler classification. Therefore, to address this, the key contributions of this study are:-
\begin{itemize}
    \item Systematic characterization of feature degradation and vulnerability under controlled noise conditions, including Additive White Gaussian Noise (AWGN), Phase noise, and combined ranges. Noise levels have been selected to represent moderate and physically plausible conditions, commonly in radar literature, AWGN ranging from $-10$ dB to $10$ dB is taken into account \cite{nguyen2024improving}. Phase noise is simulated ranging from 1-10$^\circ$. The effect of both noise types in tandem is also considered.
    \item Establishing baseline performance metrics for handcrafted features and quantifying shifts in feature importance using classifiers with fixed hyperparameters across varying noise types and intensity levels.
\end{itemize}

This systematic robustness analysis complements performance-centric investigations by providing quantitative insights into feature reliability under controlled noise. The remaining sections are organized as follows: Section 2 describes the system model and proposed methodology, Section 3 presents and discusses the results, and Section 4 concludes the paper.

\section{Proposed Methodology}

This section describes the methodology used to extract micro-Doppler features from radar signals, simulate noise perturbations, and evaluate feature robustness via classification performance degradation and importance shifts. The workflow consists of five main steps: (1) micro-Doppler spectrogram extraction, (2) Simulated noise injection, (3) feature extraction, (4) classification using Support Vector Machine (SVM) and Random Forest classifiers and (5) Analysis of feature robustness under different noise conditions using performance metrics and importance shifts.  

\subsection{Dataset and Micro-Doppler Spectrogram Extraction}
A publicly available Frequency-Modulated Continuous-Wave (FMCW) radar dataset targeting aerial objects was used in this study \cite{alexander_karlsson_2021_5845259}. Table \ref{tab:radar_specifications} contains detailed information on the radar used to collect radar returns.
The dataset contains radar returns from six different types for drones, six different species of birds and a Trihedral reflector class that represents targets with little or no micro-Doppler variations. We couple these sub classes into three super classes; drones, birds and reflectors. The dataset contains a total of 130 radar measurements. Each measurement is divided into shorter individual temporal windows, referred to as \textit{segments}. This imbalance introduces a potential confounding factor, as features derived from temporal marginals may implicitly encode observation duration rather than micro-Doppler structure. To handle this imbalance, all calculated spectrograms were cropped at center using a fixed slow-time window of length $L$, ensuring temporal standardization between classes. $L$ was selected as 32 with the variability of the data considered. Any values that fell short of this window were truncated in order to standardize the temporal axis. The spectrograms are treated as three-dimensional arrays containing range bins, slow-time samples, and consecutive segments. For each segment, the range bin with maximum accumulated energy across slow-time samples was selected to represent the target of interest. The energy for each range bin, depicted as $E[r]$, was calculated using, 

\begin{equation}
E[r]=\sum_{n=1}^{N} |x[r,n]|^{2}
    \label{eq:Energy_Range}
\end{equation}

where $x[r,n]$ is the complex radar signal at range bin $r$ and slow-time sample $n$, and $N$ is the total number of slow-time samples. This reduced the data to a single slow-time signal per segment. For each measurement, segment-level Doppler spectra were computed and concatenated along the temporal axis to form a measurement-level micro-Doppler spectrogram. 
\begin{table}[t]
\centering
\caption{Radar data specifications}
\label{tab:radar_specifications}
\begin{tabular}{ll}
\toprule
Parameter & Value \\
\midrule
Type & SAAB SIRS 1600 (FMCW) \\
Center frequency & 77 GHz \\
Bandwidth & 160 MHz \\
Range resolution & 1 m \\
Azimuth beamwidth & $1^\circ$ \\
PRF & 17 kHz \\
Scan rate & 10 Hz (mechanical) \\
Field of view & $\pm 9^\circ$ \\
Output power & 10 mW \\
\bottomrule
\end{tabular}
\end{table}

\begin{table}[t]
\centering
\caption{Measurement duration for drones, birds and reflectors revealing significant variability}
\label{tab:duration_specs}
\begin{tabular}{lccc}
\toprule
Statistic & Drone &    Bird & Reflector \\
\midrule
Mean &    146.68  &  31.88 &   17.23 \\
Standard Deviation & 108.67  &  40.75 &    9.07 \\
Median &  123.80  &  13.94 &   11.70 \\
25th percentile &  56.35   &  7.90 &  9.30 \\
50th percentile & 123.80   & 13.94 & 11.70 \\
75th percentile & 242.35   & 40.47 & 29.47 \\
\bottomrule
\end{tabular}
\end{table}

\begin{figure}[ht] 
  \begin{subfigure}[b]{0.5\linewidth}
    \centering
    \includegraphics[width=\linewidth]{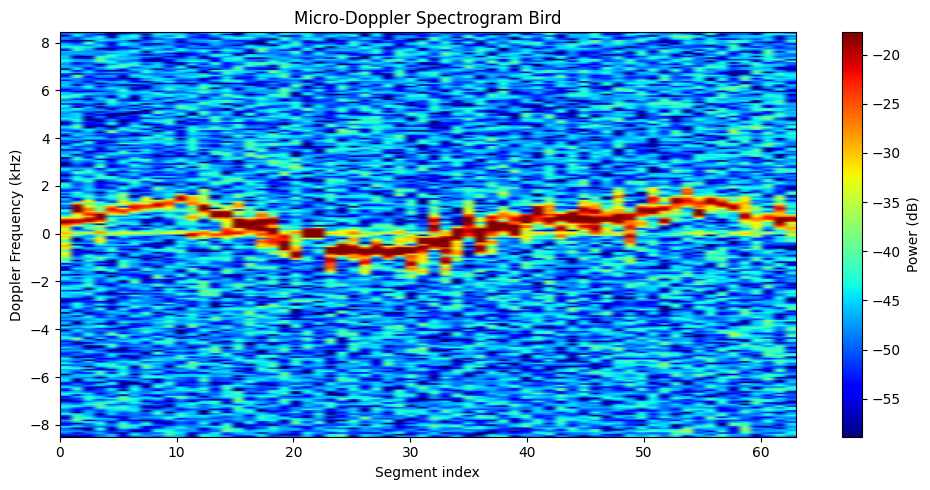} 
    \caption{Bird} 
    \label{fig:example_spectrograms:a} 
    \vspace{2ex}
  \end{subfigure}
  \begin{subfigure}[b]{0.5\linewidth}
    \centering
    \includegraphics[width=\linewidth]{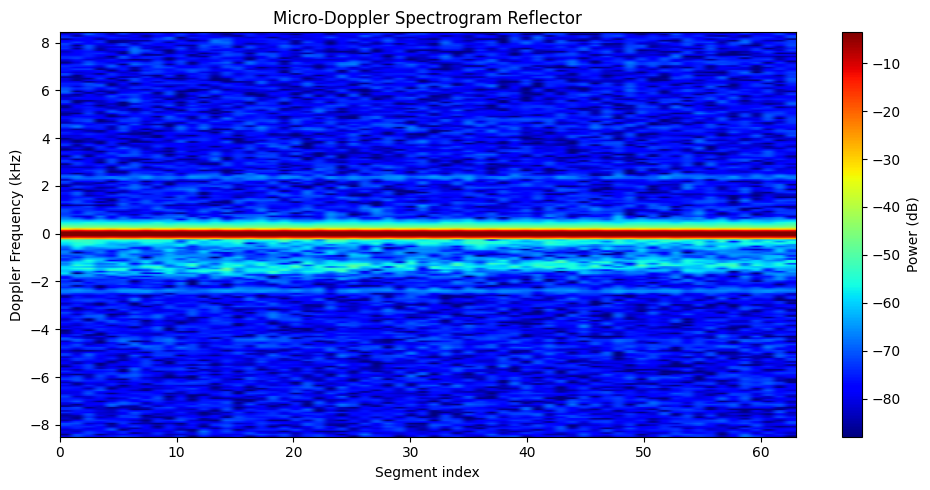} 
    \caption{Reflector} 
    \label{fig:example_spectrograms:b} 
    \vspace{2ex}
  \end{subfigure} 
  \begin{subfigure}[b]{0.5\linewidth}
    \centering
    \includegraphics[width=\linewidth]{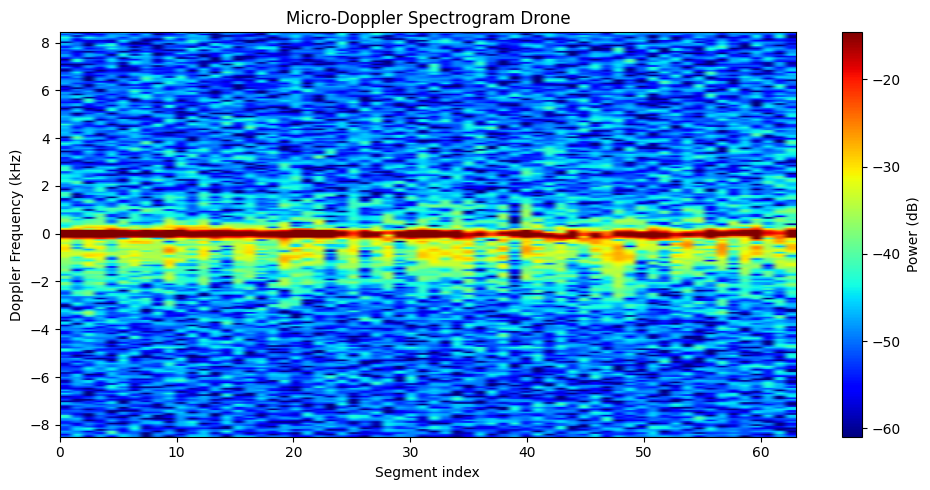} 
    \caption{Hovering drone} 
    \label{fig:example_spectrograms:c} 
  \end{subfigure}
  \begin{subfigure}[b]{0.5\linewidth}
    \centering
    \includegraphics[width=\linewidth]{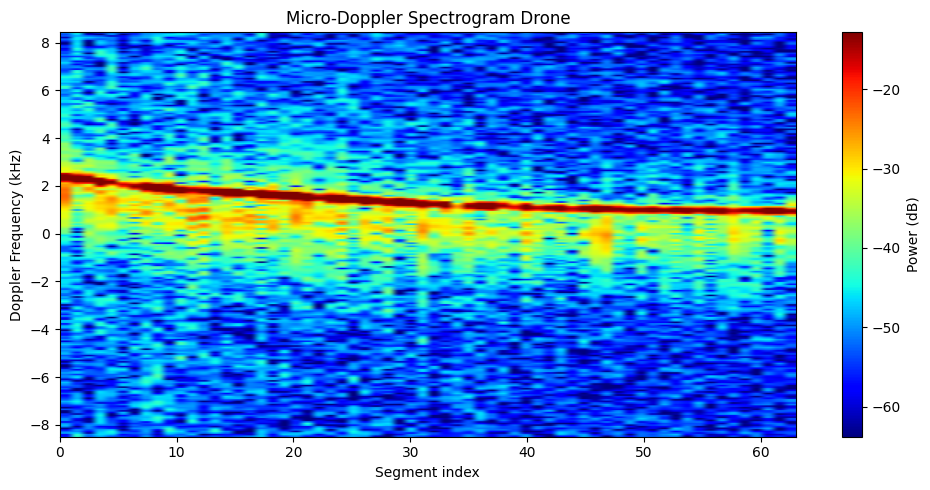} 
    \caption{Moving drone} 
    \label{fig:example_spectrograms:d} 
  \end{subfigure} 
  \caption{Example Micro-Doppler Spectrograms calculated from raw radar returns}
  \label{fig:example_spectrograms} 
\end{figure}

To convert these slow-time signals into meaningful Doppler spectrograms, several steps were taken. The mean of the signal is subtracted to cater for the DC offset that manifests at 0 Hz. The resulting signal is then multiplied with a Hanning window function to reduce spectral leakage. Fourier transform is applied to the resulting signal to obtain Doppler spectrum. The signal is then shifted using the FFT shift operation to ensure zero-Doppler is at the center. To calculate the spectrogram for each segment, the magnitude of the coefficients of the Fourier transform are squared, converted into decibels, and normalized, using 
\begin{equation}
    S(f,t) = 10 \log_{10} \left( \frac{|X[f,t]|^2}{\text{max}_{f,t}|X[f,t]|^2} \right)
    \label{eq:spectrogram}
\end{equation}
where $t$ represents time. Segment-level spectrograms were concatenated into 44 drone spectrograms, 56 birds spectrograms and 19 reflector spectrograms. Figure \ref{fig:example_spectrograms} shows examples of the spectrograms resulting from the methodology. Figure \ref{fig:example_spectrograms:a} shows the micro-Doppler effect caused by a bird flapping its wings, while Figure \ref{fig:example_spectrograms:c} and Figure \ref{fig:example_spectrograms:d} show the side lobes in the spectrograms caused by drone rotors. Figure \ref{fig:example_spectrograms:b} shows no side lobe data as expected from a reflector.

\subsection{Simulated Noise Injection}
To evaluate feature robustness, the signals were subjected to different types of simulated noise, three types of scenarios were chosen; AWGN, Phase noise, and the effect of both combined. Noise was applied to the raw signals and the resulting signals were saved separately before the spectrograms were generated, ensuring that noise effects propagate through the full time–frequency processing chain. AWGN was used to simulate thermal noise and other random background noise that the signal faces in real radar scenarios. The original signal, $x[n]$, was added to a signal, $w[n]$, that was sampled from a normal Gaussian distribution, which is normalized by the target SNR using,
\begin{equation} 
\label{eq:x_awgn}
x_{\text{awgn}}[n] = x[n] + w[n], \quad
w[n] \sim \mathcal{N}(0, \sigma^2), \quad
\sigma^2 = 10^{-\mathrm{SNR}/10} P_{\text{signal}}
\end{equation}
where SNR is in dB. $P_\text{signal}$ is calculated for each segment in the original measurements.

Raw samples were subjected to different noise levels, including −10 dB and −7 dB to simulate extremely noisy conditions, followed by a range of −5 dB to 5 dB to represent moderate corruption. 7 and 10 dB were also included to provide symmetry around SNR levels. Oscillator jitter is inherent to all radar systems and it raises the noise floor against the signal power \cite{kannanthara2023whole}. The signals were  subjected to phase noise with independent and identically distributed phase perturbations of 1$^\circ$ – 10$^\circ$ (converted to radians). To apply phase noise, $e^{j\phi}$ was multiplied to $x[n]$ as shown in Eq \ref{eq:x_phase},

\begin{equation} 
x_{\phi\ noise}[n] = x[n] \cdot e^{j\phi[n]}, \quad  \phi[n] \sim \mathcal{N}(0, \sigma ^ 2).
\label{eq:x_phase} 
\end{equation}

The signal was also subjected to a combined effect of both AWGN and phase noise. This application was done in a tiered fashion; mild, moderate, and severe levels, resulting in nine distinct levels. Table~\ref{tab:noise_modes_compact} shows a list of all noise types and tiers included in the experiments.

\begin{table}[t]
\centering
\caption{Summary of noise modes applied to radar signals}
\label{tab:noise_modes_compact}
\begin{tabular}{l l l l}
\toprule
\textbf{Noise Mode} & \textbf{AWGN (dB)} & \textbf{Phase ($^\circ$)} & \textbf{Severity} \\
\midrule
Raw                 & —      & —   & None     \\
\midrule
\multirow{3}{*}{AWGN only}
                    & $-10$ to $-7$  & —   & Severe   \\
                  & $-5$ to $0$    & —   & Moderate \\
                  & $5$ to $10$    & —   & Mild     \\
\midrule
\multirow{3}{*}{Phase only}
                    & —      & $1$–$3$   & Mild     \\
                  & —      & $4$–$7$   & Moderate \\
                  & —      & $8$–$10$  & Severe   \\
\midrule
\multirow{9}{*}{Combined}
                    & $-3$   & $1$   & Mild     \\
                  & $-2$   & $2$   & Mild     \\
                  & $-1$   & $3$   & Moderate \\
                  & $0$    & $4$   & Moderate \\
                  & $1$    & $5$   & Moderate \\
                  & $7$    & $1$   & Moderate \\
                  & $2$    & $6$   & Severe   \\
                  & $3$    & $7$   & Severe   \\
                  & $5$    & $8$   & Severe   \\
\bottomrule
\end{tabular}
\end{table}

\subsection{Feature Extraction}
Features were extracted from the calculated spectrograms. These features were selected based on their physical interpretations. Since the motion physics of the spectrograms encode different characteristics, features capturing these differences are extracted to classify.  Micro-Doppler effect caused by drones has a higher frequency since the parts causing the effect move at a higher frequency, whereas for birds this effect spreads over more time bins due to their slower flaps. Hovering drones tend to be centered around the zero Doppler while birds need to move more to stay in flight, thus their spectrograms usually are more distributed. Valuable discriminative features that capture this information were calculated using frequency and time marginals. Zero-padded segments introduced for temporal standardization were automatically detected via an energy-based threshold and excluded from marginal computation. This ensures that duration normalization does not bias entropy or variance-based descriptors The spectrograms were converted from its logarithmic representation into its linear representation by using the following relationship,
\begin{equation}
    P(f_i,t_j) = 10 ^ {\frac{S(f_i,t_j)}{10}}    ,
\end{equation}
which gives the linear power representation of the spectrograms. Frequency and time marginals were calculated by,
\begin{equation}
\label{eq:marginals}
P_f(f_i) = \sum_{j=1}^{T} P(f_i, t_j), \quad
P_t(t_j) = \sum_{i=1}^{F} P(f_i, t_j)
\end{equation}
\noindent and the total power of the signal is calculated using,
\begin{equation}
P_{tot} = \sum_{i=1}^{F} P_f(f_i).
\end{equation}

The normalized distributions of the frequency and time marginals are given by:
\begin{equation}
\label{eq:nomralized_marginals}
\tilde{P}_f(f_i) = \frac{P_f(f_i)}{P_{tot}}, \quad
\tilde{P}_t(t_j) = \frac{P_t(t_j)}{P_{tot}}
\end{equation}

Using the full spectrogram and the marginals in Eq \ref{eq:marginals}, temporal and spectral features were calculated. The normalized marginals in Eq \ref{eq:nomralized_marginals} are treated as discrete probability mass functions for entropy and statistical moment calculations. These represent the temporal and spectral dispersion of micro-Doppler energy and variability in Doppler information across temporal and spectral axes. The dispersion represents the difference in the periodic motion of rotors in drones against the relatively irregular pattern of wing flaps for birds.

Third, Doppler distribution features such as Doppler centroid, Doppler spread, and higher-order statistical descriptors were also extracted. These represent the dominant Doppler content and its variability. The statistical features capture the asymmetry and peakedness in the distributions of the marginals. The squared magnitude spectrogram coefficients are treated as proportional to signal energy within each time–frequency bin, and all energy-based features are computed from this representation. The extracted features include Side lobe energy ratio, Side lobe entropy, Spectral entropy, Temporal energy variance, Temporal entropy, 80th percentile of Doppler bandwidth, Doppler spread, Zero Doppler ratio, Skewness, and Kurtosis. These features are defined as follows:
\paragraph{Side lobe Features}

The side lobe energy ratio is the energy of the total spectrogram except a selected percentage of the bins centered at zero Doppler. Let $C$ denote the set of Doppler bins within the central $\alpha = 15\%$ bandwidth around zero Doppler. The side lobe bins are implicitly the complement of $C$.

The Side lobe Energy Ratio (SLER) is then defined as

\begin{equation}
    \text{SLER} = \frac{P_{tot}- \Sigma_{i\in {C}}P_f (f_i)}{P_{tot}}
\end{equation}

While, the side lobe entropy is calculated using the entropy of the normalized side lobe energy distribution.

\paragraph{Temporal Features}

To capture temporal irregularities, temporal entropy was calculated from normalized time marginals. Temporal energy variance is computed on the unnormalized time marginal to preserve absolute energy fluctuations. Let 
\begin{equation}
    \mu_t = \frac{1}{T} \sum_{j=1}^{T}P_t(t_j)
\end{equation}
then temporal energy variance is given by,
\begin{equation}
    \sigma^2_t = \frac{1}{T} \sum_{j=1}^{T}(P_t(t_j) - \mu_t)^2
\end{equation}

\paragraph{Doppler Features}
Specific Doppler based features were extracted to capture Doppler based information from the spectrograms. The $\text{80}^{th}$ percentile of the cumulative Doppler distribution is calculated to measure the effective Doppler extent and reduce sensitivity to outliers.  The energy ratio of a small window around zero Doppler is used to capture dominant stationary or near-stationary components such as reflectors. Another feature is the Doppler spread, it is the power-weighted standard deviation of the Doppler frequency distribution. 

Higher-order statistical moments such as skewness and kurtosis are calculated for the frequency marginal. In total, a fixed set of ten hand-crafted features was extracted per segment and used as input to the classification models described in the following subsection. 

\subsection{Classification}
To ensure that these features held sufficient information for class separability, SVM and Random Forest classifiers were trained on the extracted features. SVM and Random Forest were trained on different combinations of the feature sets and the best feature set was chosen. An SVM was also trained on the individual features to analyze their standalone classification performance. For sanity checks, 20\% of the data was considered a hold out set, this data was not used for training any of the SVM or Random Forest models. This holdout set was later used to construct a confusion matrix. The matrix represents the actual classes against the predicted labels which helps understand if the model is correctly classifying the classes. The remaining 80\% of the data is used for training, using 5-fold stratified cross validation. Stratified cross validation was done to reduce bias and keep overall data distribution, since reflector class is under represented. Cross-validation splitting was performed at the measurement level to prevent leakage between temporally adjacent segments. Prior to classification, the feature data was preprocessed by standardizing each vector using a z-score normalization ensuring a zero centered mean and one unit variance. To convert semantic labels, a label encoder was used to map the class labels to integers indices for each class.  

\begin{figure}[!t]
  \centering
  \includegraphics[width=\textwidth]{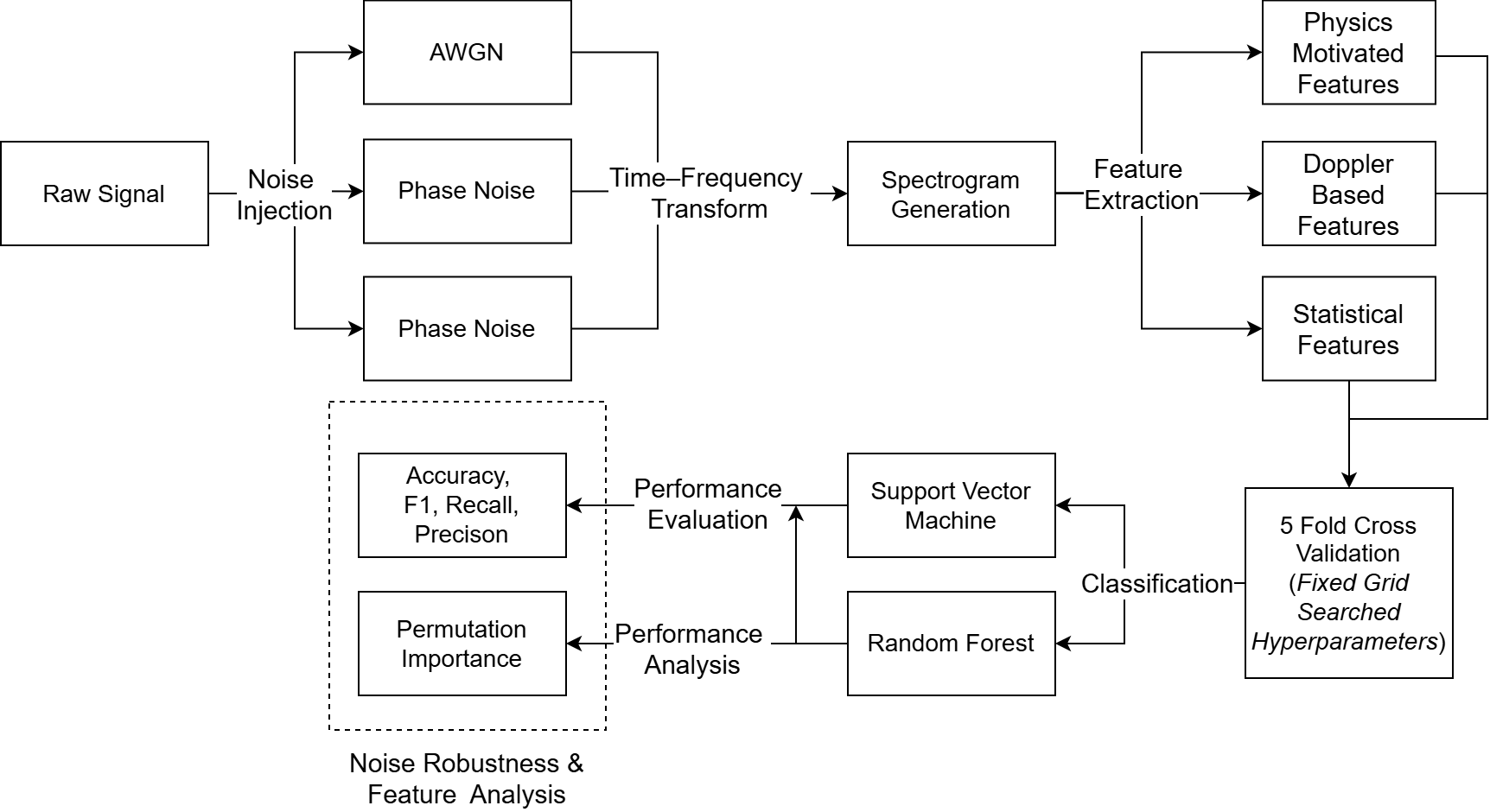}
  \caption{High level system diagram for proposed methodology}
  \label{fig:methodology}
\end{figure}

SVM and Random Forest were specifically chosen for their characteristics. SVM was chosen for its ability of clear margin maximization with high-dimensional spaces, while Random Forest was chosen for its non-linear boundaries and interpretability of feature interactions. A grid search was used to choose the best hyper-parameters for raw data. The search was optimized for macro-F1 scores and these parameters were then kept constant for all noise modes ensuring equal comparability. 
To ensure reproducibility, a random seed was chosen with a value of 42 which was used by all classifiers. For the single feature classification SVM used a linear kernel  and $C$ = 1.0. For multi-feature classification, the SVM used a Radial Basis Function (RBF) kernel, $C$ = 100 and $\gamma$ = 0.1, allowing non linear combinations to be learned.
Finally, an Random Forest classifier was used to calculate permutation importance to interpret how shifts in feature relevance. Feature scaling was not applied to the Random Forest, as tree-based models are invariant to monotonic feature transformations. The Random Forest classifier used $300$ estimators with a max depth of $5$.

To analyze model performance, accuracy, precision, recall and F1 scores were calculated as metrics for all setups. Performance metrics were computed per fold and averaged across folds. This protocol ensures that variations in performance can only be attributed to noise effects rather than differences in training or evaluation procedures. The permutation importance was collected to understand which features hold more power for separability under specific noise conditions. This approach was chosen over impurity-based importance to avoid bias toward high variance features and to directly measure performance degradation when feature information is disrupted.
All data processing and analysis were performed using Python 3.11 with libraries including numpy and pandas for data manipulation, scikit-learn for machine learning classifiers, cross-validation, and permutation importance, and matplotlib for visualization.
As shown in Figure \ref{fig:methodology}, the raw radar signals were first corrupted according to the specified noise model prior to spectrogram generation. Features were then extracted, partitioned into 5-fold stratified cross validation folds from the resulting spectrograms and then used as inputs to the classifiers.
Analysis on noise performance was performed by evaluating these metrics on corrupted signals. This analysis was used to assess feature robustness and relevance rather than to optimize for higher classification accuracy.
An ablation study was then conducted with adding and removing features to the proposed feature set. Doppler based features and higher-order statistical features were tested for all noise levels and metrics were calculated and compared against the proposed set. All hyperparameters were fixed across experiments to isolate the effect of feature selection from model tuning.

\section{Results and Discussion}
This section evaluates the performance and robustness of the physics-motivated features extracted from the micro-Doppler spectrograms under different noise conditions. Classification metrics are analyzed for classifiers over all noise conditions to understand performance degradation. From the initial ten handcrafted features a subset was selected comprising of Side lobe energy ratio, Side lobe entropy, Spectral entropy, Temporal energy variance, and Temporal entropy. Selection was guided by measurement-level cross-validated performance and consistent robustness trends under controlled noise perturbations. Features that exhibited limited discriminative contribution or high sensitivity to noise were excluded from the proposed feature set to avoid redundancy. Importantly, the selected feature set is systematically compared against the full feature set and alternative combinations, and the behavior is further examined through Random Forest permutation importance analysis.

\subsection{Baseline Classification Performance}

\begin{figure}[!ht]
\centering
\begin{subfigure}{.5\textwidth}
  \centering
  \includegraphics[width=0.80\textwidth]{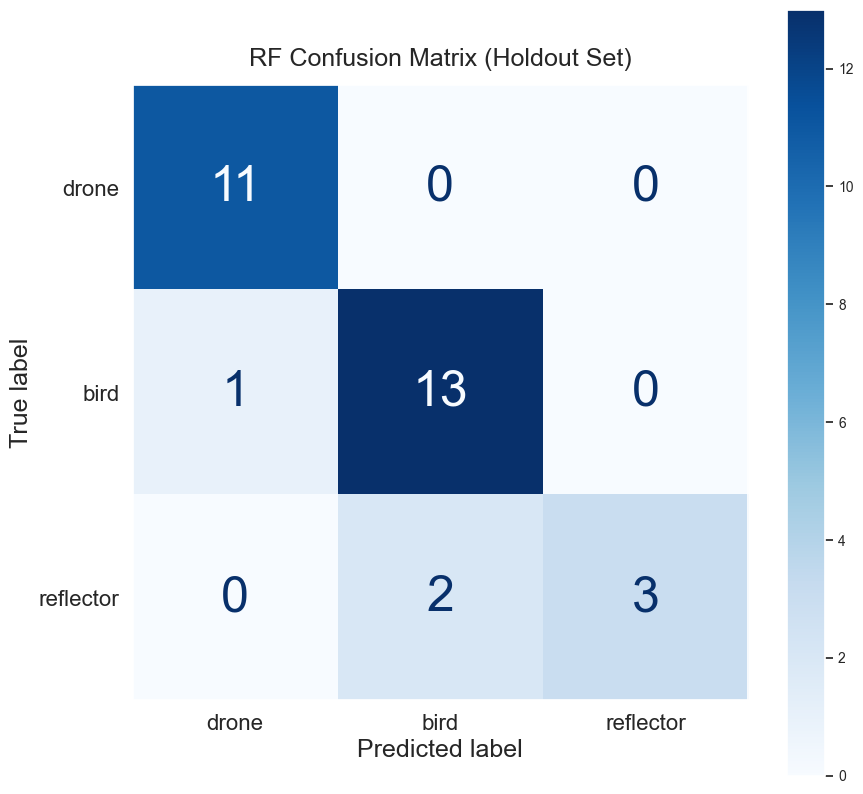}
  \caption{Confusion matrix for Random Forest}
  \label{fig:sub1}
\end{subfigure}%
\begin{subfigure}{.5\textwidth}
  \centering
  \includegraphics[width=0.80\textwidth]{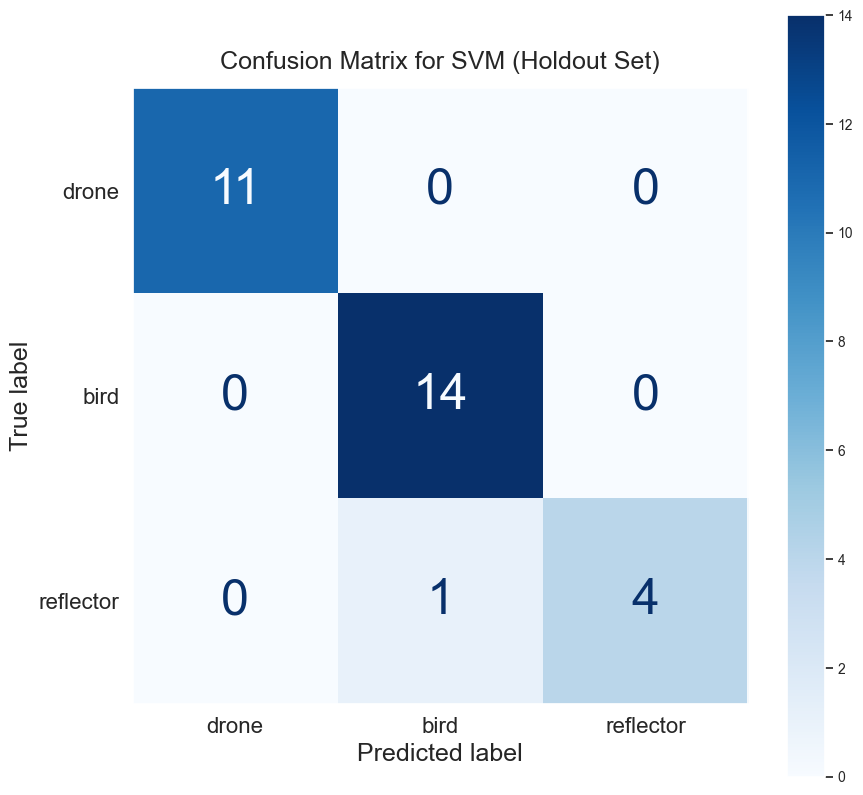}
  \caption{Confusion matrix for SVM}
  \label{fig:sub2}
\end{subfigure}
\caption{Confusion matrices with results of each classifier on the holdout set using raw data}
\label{fig:cm}
\end{figure}

The baseline performance of the classifiers was evaluated using the proposed methodology under noise-free conditions. Table \ref{tab:svm_baseline_metrics} shows metrics for raw samples using SVM and Random Forest utilizing the proposed feature set. SVM achieved an accuracy of 0.916 $\pm$ 0.095, while Random Forest achieved an accuracy of 0.916 $\pm$ 0.060 for raw samples. In addition, macro-averaged F1, Precision and Recall are also reported, to show class variability. Recall is particularly important due to the low number of samples for the reflector class. For macro averaged F1, SVM scored $0.909$, while Random Forest scored $0.912$ indicating balanced class discrimination. Minor variations across metrics reflects differences in class-specific separability, particularly for the underrepresented reflector class. Figure \ref{fig:cm} shows confusion matrices on a holdout set that was held out of the training set before cross-validation. This is done as a sanity check to ensure that the model is able to retain classification ability for completely unseen data.
\begin{table}[t]
\centering
\caption{Baseline performance of Random Forest and multi-feature SVM on raw samples}
\label{tab:svm_baseline_metrics}
\begin{tabular}{lcc}
\toprule
Metric & SVM & Random Forest \\
\midrule
Accuracy ($\mu \pm \sigma$) & 0.916 $\pm$ 0.095 & 0.916 $\pm$ 0.060 \\
F1 ($\mu$) & 0.909 &  0.912 \\
Precision ($\mu$) & 0.950 &  0.947 \\
Recall ($\mu$) & 0.901 &   0.892 \\
\bottomrule
\end{tabular}
\end{table}

These results act as a reference point showing that the proposed features have the ability to discriminate between the classes without noise injection. This prompts an in depth robustness analysis for these features under controlled simulated noise.

\subsection{SVM for individual features}
\begin{figure}[ht]
\centering
\begin{subfigure}{0.48\textwidth}
  \centering
  \includegraphics[width=\textwidth]{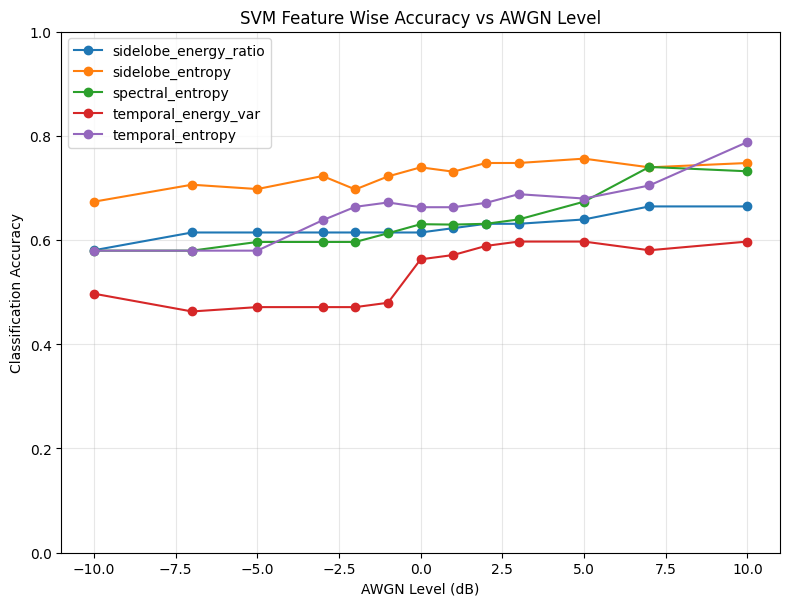}
  \caption{AWGN}
  \label{fig:svm_feature_awgn}
\end{subfigure}
\hfill
\begin{subfigure}{0.48\textwidth}
  \centering
  \includegraphics[width=\textwidth]{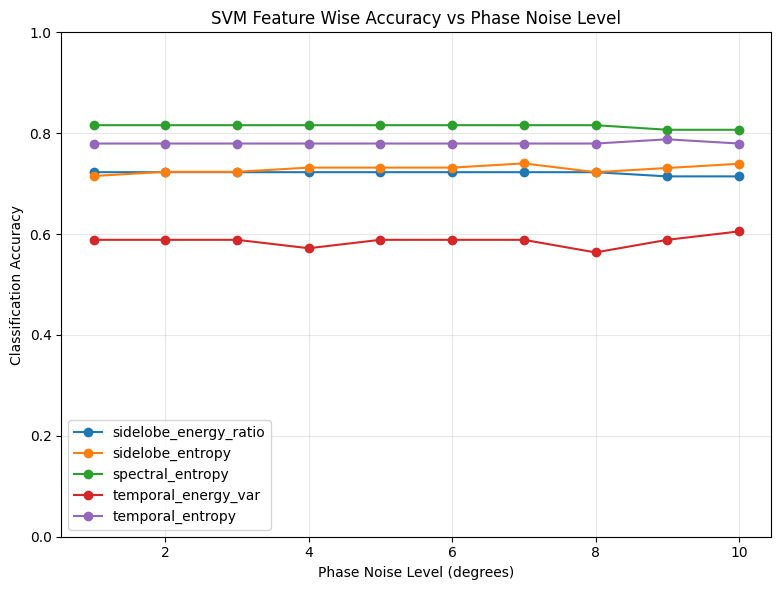}
  \caption{Phase Noise}
  \label{fig:svm_feature_phase}
\end{subfigure}

\vspace{0.3cm}

\begin{subfigure}{0.48\textwidth}
  \centering
  \includegraphics[width=\textwidth]{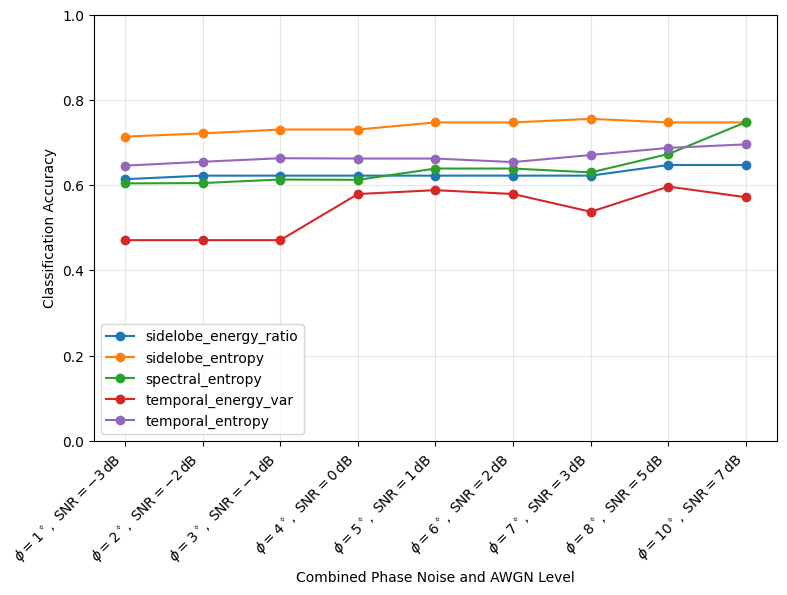}
  \caption{Combined Noise}
  \label{fig:svm_feature_combined}
\end{subfigure}

\caption{Feature-wise SVM classification metrics under different noise models as a function of noise level.}
\label{fig:svm_feature_noise}
\end{figure}

Analysis of individual features was conducted by using raw and noisy samples resulting in accuracies demonstrated in Figure \ref{fig:svm_feature_noise}.

Under AWGN, all features show incremental increase in separability as signal quality improves. Side lobe energy ratio and temporal entropy demonstrate substantial discriminative ability across all noise levels. Accuracy for side lobe entropy increases as the SNR improves to 0 dB, showing sensitivity to AWGN noise but degrading gracefully under noise while avoiding collapse. This could be attributed to the rise in the noise floor, since side lobe entropy takes the full spectral distribution into account and it becomes more uniform due to noise spreading energy across frequency bins, which increases entropy and reduces discriminative contrast. Spectral entropy and temporal energy variance show a similar trend of increase in accuracy as SNR improves but have a lower overall accuracy. Side lobe Energy Ratio remains resistant to noise for most ranges showing little fluctuations. 

Under phase noise, Spectral entropy remains the most discriminative, consistently offering approximately 82\% accuracy even under extreme noise. Entropy based features remain consistent across all levels, where side lobe entropy levels climb up to 78\% above 8$^\circ$ noise. Temporal energy variation suffers under higher levels of phase noise dropping to a low 58\%. Phase noise perturbs phase coherence, which redistributes energy in the Doppler domain and alters side-lobe structure. This redistribution may amplify differences in side lobe structure, partially explaining the observed robustness of side lobe energy ratio. When subjected to a combination of both noise types, overall accuracy is slightly lower as expected. Entropy based features such as temporal entropy, side lobe entropy and side lobe entropy remain consistently distinguishable for the model all being above 75\% accuracy as signal quality improves. Side lobe energy ratio still remain at lower than the rest but do not degrade drastically even under severe noise showing some level of robustness. Temporal energy variance drops significantly under severe AWGN showing that noise disrupts the global spread of energy by raising the noise floor effecting the spread of noise overall in the spectrogram that is captured by this feature. These results show that each of the proposed feature set carries non-trivial classification ability even under different types of noise. This lays down the foundation for a deeper analysis of the performance of these features in tandem.

\subsection{Multi-feature SVM under noise}
\begin{figure}[!t]
\centering
\begin{subfigure}{0.48\textwidth}
  \centering
  \includegraphics[width=\textwidth]{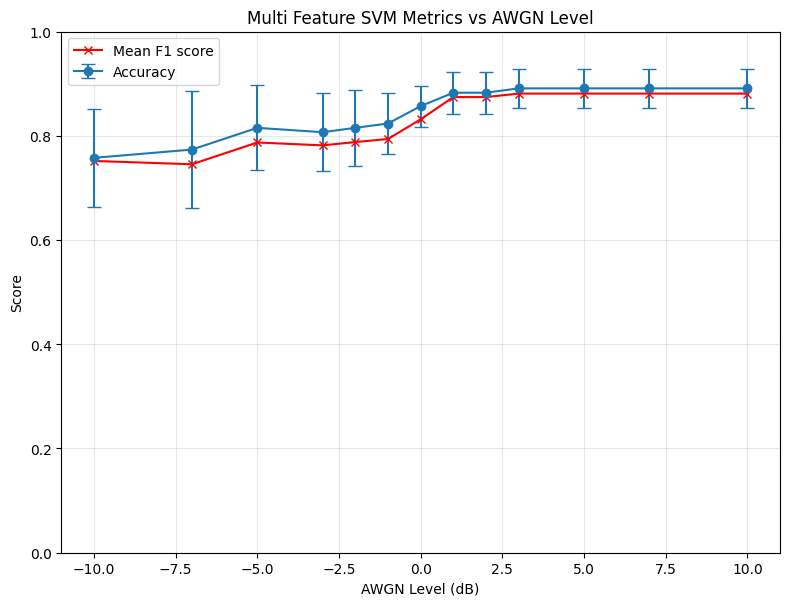}
  \caption{AWGN}
  \label{fig:multi_svm_f1_awgn}
\end{subfigure}
\hfill
\begin{subfigure}{0.48\textwidth}
  \centering
  \includegraphics[width=\textwidth]{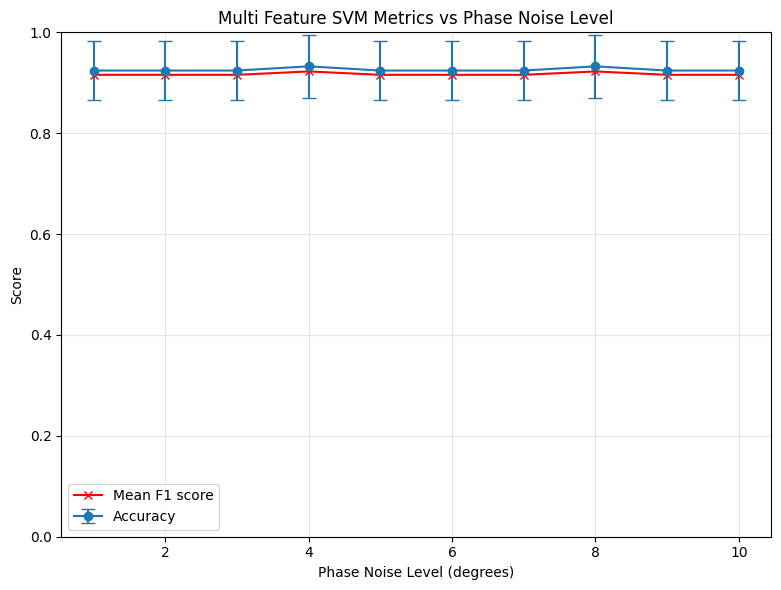}
  \caption{Phase Noise}
  \label{fig:multi_svm_f1_phase}
\end{subfigure}
\vspace{0.3cm}

\begin{subfigure}{0.48\textwidth}
  \centering
  \includegraphics[width=\textwidth]{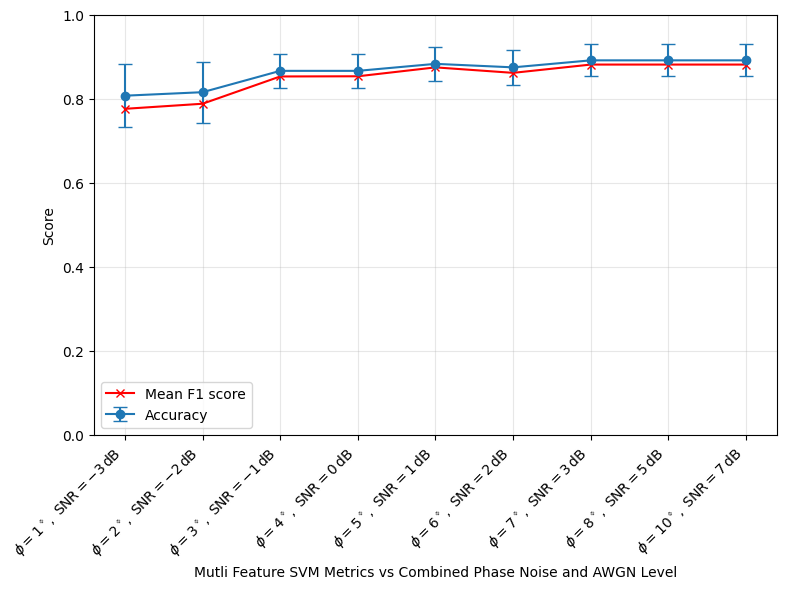}
  \caption{Combined Noise}
  \label{fig:multi_svm_f1_combined}
\end{subfigure}

\caption{Multi-feature SVM classification metrics under different noise models as a function of noise level.} 
\label{fig:multi_svm_metrics}
\end{figure}

The results in Figure~\ref{fig:multi_svm_metrics} show the classification accuracy and F1 score of the multi-feature SVM. The SVM used selected feature set and was subjected to AWGN, phase noise, and different combinations the ability of the classifier under degraded signal quality. For clean data, the multi-feature SVM yielded an accuracy of 0.916 $\pm$ 0.095 with an F1 score of 0.909. Figure~\ref{fig:multi_svm_f1_awgn} demonstrates that under AWGN the features degrade gracefully even when subjected to extreme noise below -5 dB with F1 score falling to $0.773$ and accuracy reduced to 0.783 $\pm$ 0.092 at -10 dB. For a wide range of noise levels, the model remains relatively stable. Improvements can be noted for values above 1 dB, however, further enhancing SNR leads to stagnation in accuracy, indicating that discriminative ability gets saturated once sufficient signal quality improves. At 10 dB, the accuracy and F1 improves to $0.891 \pm 0.038$ and $0.881$, respectively. The proposed feature set retains discriminative power even when amplitude is subjected to substantial corruption. 

Moreover, Figure \ref{fig:multi_svm_f1_phase} shows that the selected features seem to be fairly resistant to phase noise, showing a drop in accuracy to $0.908 \pm 0.045$ and F1 score to $0.892$ only when subjected to severe noise of $10^\circ$. Phase noise seems to have minimal effect on the accuracy of the feature set pointing to the fact that the proposed features are relatively phase invariant up to these noise levels. When subjected to a combination of both, the SVM still stays relatively stable in classification accuracy and F1 but follows the same trend as AWGN showing that as the signal content recovers so do accuracy and F1 metrics. For further in depth feature-wise behavior and the effect of noise on these features, a Random forest classifier was trained and permutation importance was analyzed. 

\subsection{Random Forest classification and Permutation-Based Feature Importance Analysis}
\begin{figure}[!t]
\centering
\begin{subfigure}{0.48\textwidth}
  \centering
  \includegraphics[width=\textwidth]{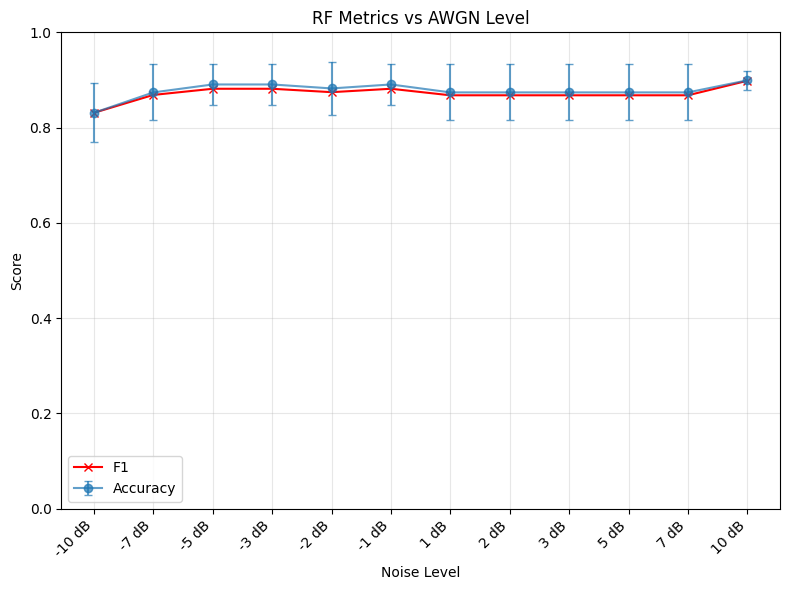}
  \caption{AWGN}
  \label{fig:rf_accuracy_awgn}
\end{subfigure}
\hfill
\begin{subfigure}{0.48\textwidth}
  \centering
  \includegraphics[width=\textwidth]{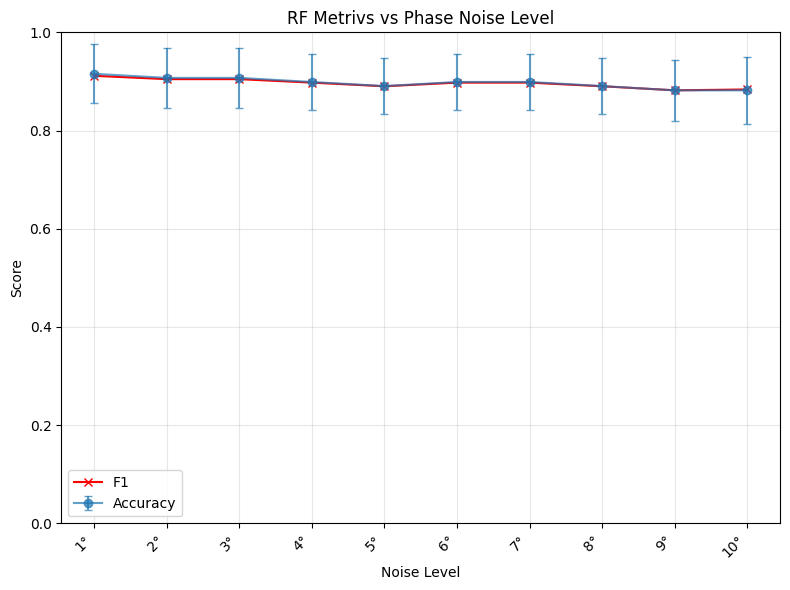}
  \caption{Phase Noise}
  \label{fig:rf_accuracy_phase}
\end{subfigure}
\vspace{0.3cm}
\begin{subfigure}{0.48\textwidth}
  \centering
  \includegraphics[width=\textwidth]{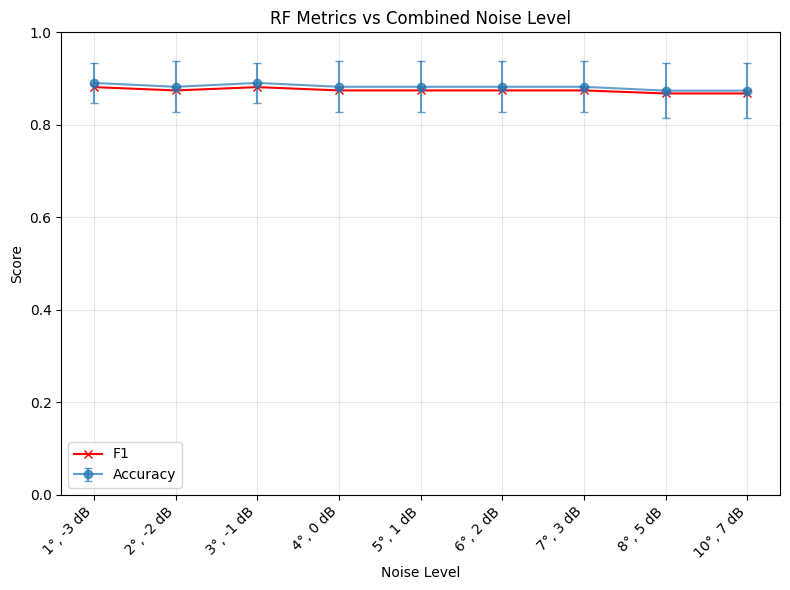}
  \caption{Combined Noise}
  \label{fig:rf_accuracy_combined}
\end{subfigure}
\caption{Random Forest classification metrics under different noise models as a function of noise level.}
\label{fig:rf_accuracy_noise}
\end{figure}

\begin{figure}[!t]
\centering
\begin{subfigure}{0.48\textwidth}
  \centering
  \includegraphics[width=\textwidth,height=0.40\textheight,keepaspectratio]{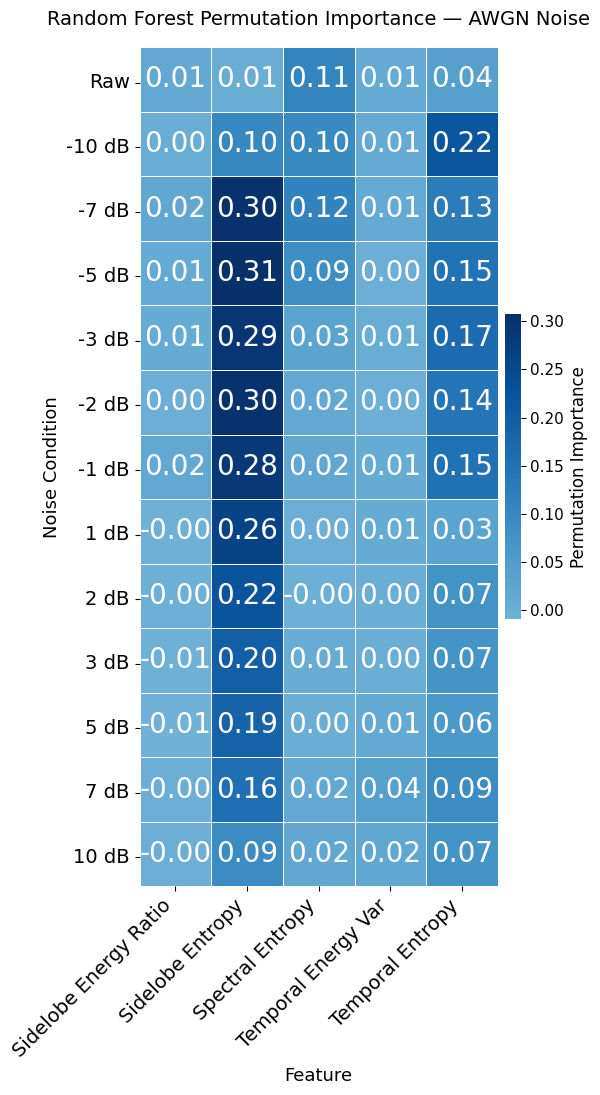}
  \caption{AWGN}
  \label{fig:rf_feature_imp_awgn}
\end{subfigure}
\begin{subfigure}{0.48\textwidth}
  \centering
  \includegraphics[width=\textwidth,height=0.40\textheight,keepaspectratio]{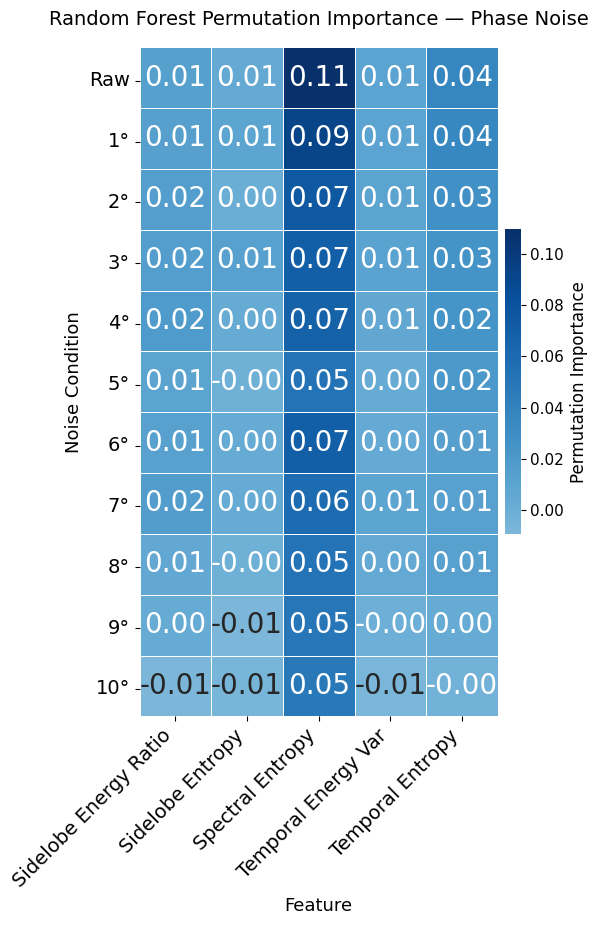}
  \caption{Phase Noise}
  \label{fig:rf_feature_imp_phase}
\end{subfigure}

\begin{subfigure}{0.48\textwidth}
  \centering
  \includegraphics[width=\textwidth,height=0.40\textheight,keepaspectratio]{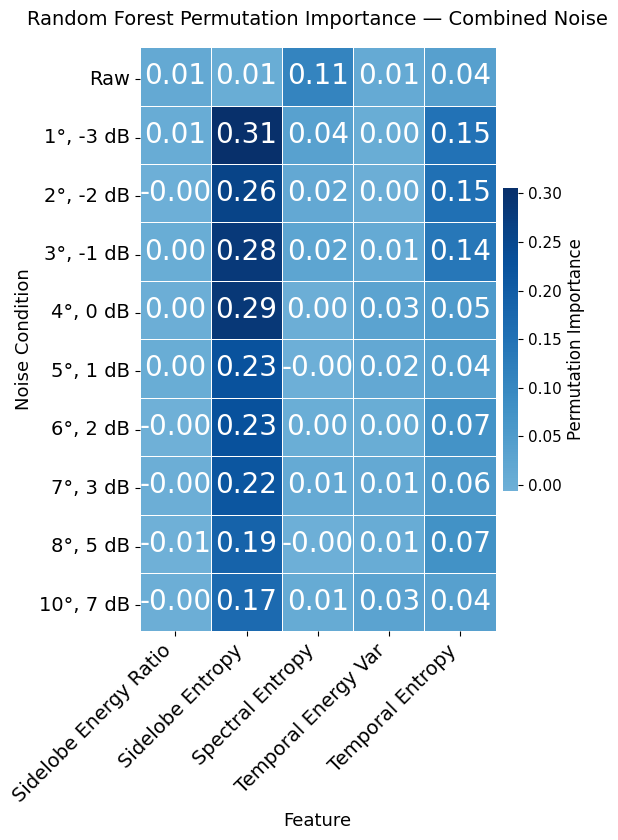}
  \caption{Combined Noise}
  \label{fig:rf_feature_imp_combined}
\end{subfigure}

\caption{Random Forest with Feature wise permutation importance as a function of noise}
\label{fig:rf_feature_importance}
\end{figure}

To analyze the feature interactions and non-linearity, a Random Forest classifier is trained on the data with noisy and clean data. The robustness of the Random Forest classifier was evaluated by progressively increasing noise levels. Figure~\ref{fig:rf_accuracy_noise} illustrates the classification accuracy of the Random forest as a function of SNR, phase noise and combined noise at different levels of noise.
Across all noise conditions, classification performance degrades gradually as signal quality weakens rather than collapsing abruptly, indicating a degree of robustness in the handcrafted feature representation. When signal quality degrades severely at -10 dB, the macro-averaged F1 score remains at $0.831$ with an accuracy of $0.831 \pm 0.062$, showing that the model remains decently stable even when signal quality degrades significantly. Under phase noise the model retains an F1 score of $0.884$ with an accuracy of $0.882 \pm 0.069$.  This is indicative of the feature sets discriminative ability with Random Forest classifier as well. To interpret what features are given importance under different noise levels, Figure \ref{fig:rf_feature_importance} shows permutation importance of each feature across different noise levels.

When no noise is present, entropy based features, temporal entropy and spectral entropy, dominate the classifier's decision. This is due to their ability to capture global dispersion of energy, drone based spectrograms produce concentrated Doppler structure while bird flapping generate irregular and distributed energy patterns. Entropy based features capture this contrast directly, thus explaining their higher baseline permutation importance.

Under AWGN, a systematic shift is observed in the importance for these features demonstrated in Figure \ref{fig:rf_feature_imp_awgn}. Side lobe entropy, temporal entropy and spectral entropy remain dominant in the classifiers decision under severe conditions while side lobe entropy becomes less important when signal quality is improved beyond -5 dB. This can be attributed to additive noise raising the overall noise floor and spreading energy more uniformly across Doppler bins. Resulting in entropy based features being relatively important because of their dependence on distributional shape rather than absolute amplitude. Side lobe energy ratio also increases in importance but only when signal quality is extremely degraded. This reflects the dependence of the feature on complementary information when signal quality degrades significantly. Temporal energy variance negatively impacts the classifiers decision under noise often having a negative permutation importance. Variance based descriptors suffer from noise due to the fluctuations it introduces. Although, the standalone importance remains low, it is observed that removing this feature reduced the overall performance. This points to the fact that some features still retain complementary information.

When the signal is corrupted by phase noise, as shown in Figure \ref{fig:rf_feature_imp_phase}, the feature importance shows the same trend as its raw counter part, where spectral entropy dominates. This corroborates the earlier observation in Figure \ref{fig:svm_feature_noise}, that the relative performance of features seems to be robust to phase changes. This indicates that the feature set has a certain resilience to phase noise.
Under the combined effect of both noise types, the importance follows the same trend as when it was subjected to AWGN. Entropy based features remain dominantly more important, while temporal energy often has low or negative importance.
The results indicate that certain features maintain consistent importance across noise levels, while others exhibit increased variability or reduced contribution under low SNR conditions. 

\subsection{Ablation Study}
An ablation study was conducted to evaluate the impact of the other features that were left out of the proposed set. The study shows that adding other features to the selected five, negatively impacts the classification performance. Table ~\ref{tab:ablation} summarizes the metrics obtained with the full set of ten features and other feature subsets for Random Forest classifier.

\begin{figure}[!t]
\centering
\begin{subfigure}{0.48\textwidth}
  \centering
  \includegraphics[width=\textwidth,height=0.40\textheight,keepaspectratio]{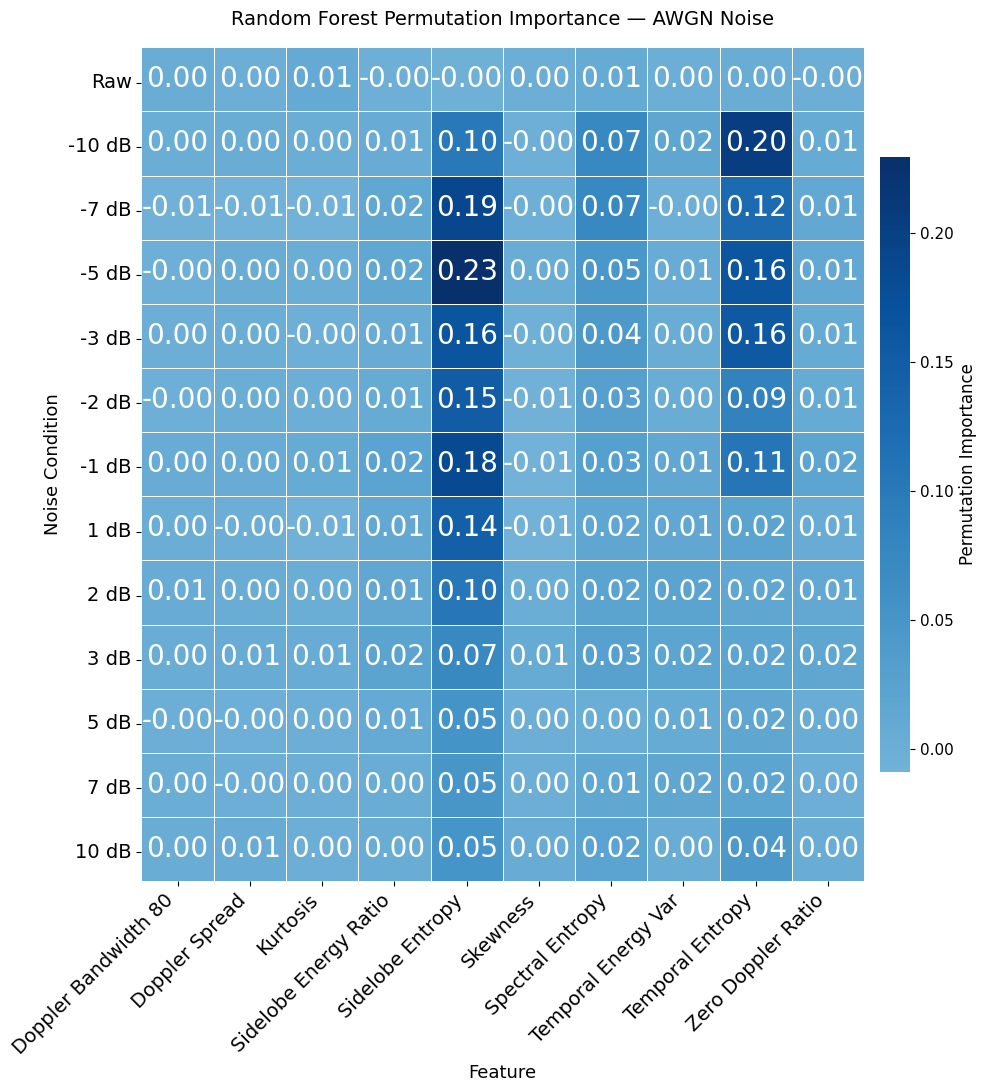}
  \caption{AWGN}
  \label{fig:rf_feature_imp_albation_awgn}
\end{subfigure}
\hfill
\begin{subfigure}{0.48\textwidth}
  \centering
  \includegraphics[width=\textwidth,height=0.40\textheight,keepaspectratio]{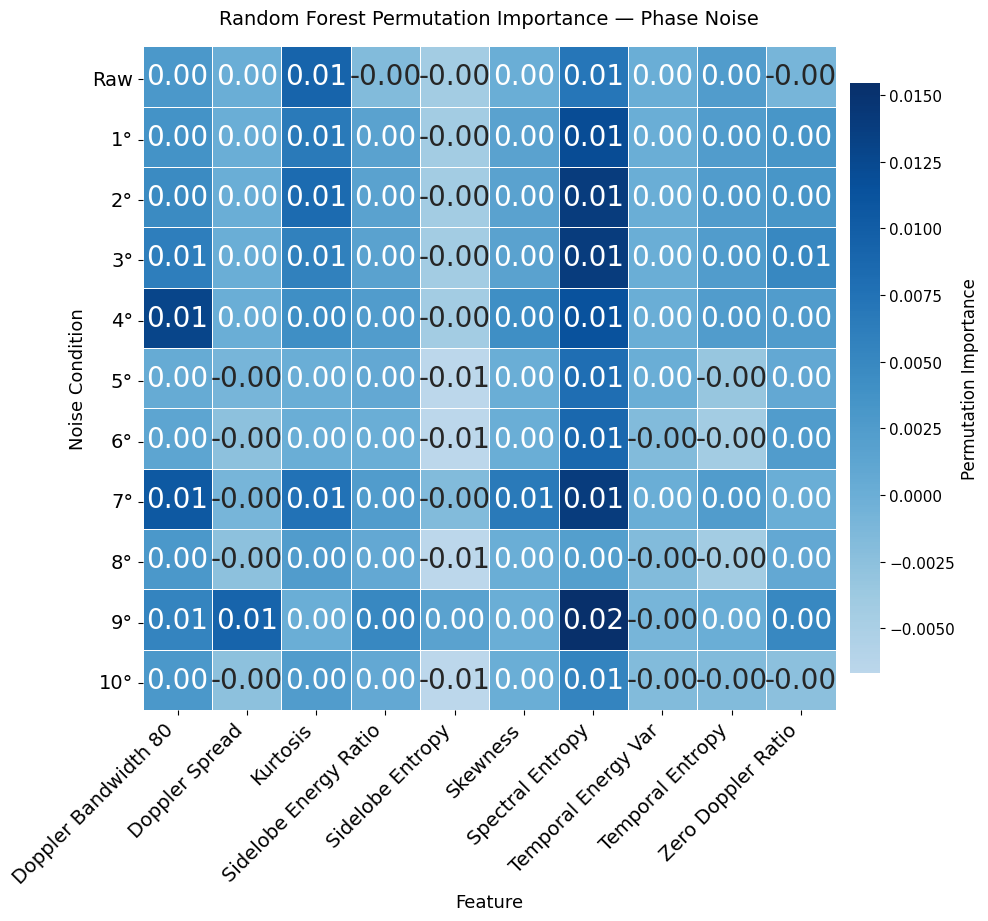}
  \caption{Phase Noise}
  \label{fig:rf_feature_imp_albation_phase}
\end{subfigure}
\vspace{0.3cm}

\begin{subfigure}{0.48\textwidth}
  \centering
  \includegraphics[width=\textwidth,height=0.40\textheight,keepaspectratio]{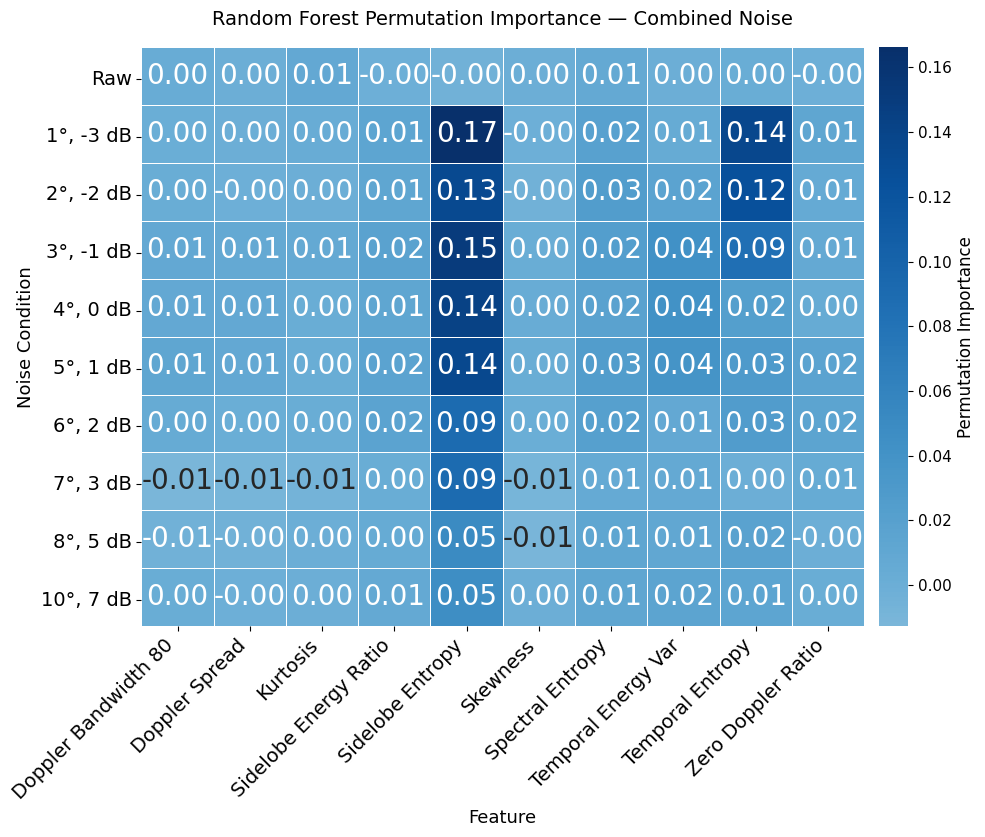}
  \caption{Combined Noise}
  \label{fig:rf_feature_imp_albation_combined}
\end{subfigure}

\caption{Random Forest with Full Feature set wise permutation importance as a function of noise}
\label{fig:rf_feature_importance_albation}
\end{figure}

The proposed feature set in Table ~\ref{tab:ablation} records the highest metrics with a mean accuracy of $0.916$ with a standard deviation of $0.060$ across folds with high recall of $0.892$. This shows that the model can successfully distinguish between the classes. 
Adding Doppler based features such as $80^{\text{th}}$ percentile of the Doppler bandwidth, Doppler spread and Zero Doppler Ratio perform worse than proposed feature set. These features show $0.890$ accuracy with $0.080$ standard deviation across folds. The macro F1 score falls considerably to $0.889$, this can be attributed to the low recall. For these features the lack of separability comes from the concentration around the zero Doppler for hovering drones and reflectors. 
Statistical features were also considered in addition to the proposed feature set, such as skewness and kurtosis. These features did not contribute significantly and the metrics are comparable to Doppler-based features. This comes from the redundancy between these features, for example, Doppler spread can be mathematically understood as the second order moment since it measures the distribution of the Doppler bandwidth. Statistical features depend on specific spectrum based features which makes them sensitive to noise and thus they lack generalization. Their partial correlation comes from their dependence on spectral energy distribution. This is also observed in the permutation importance map shown in Figure \ref{fig:rf_feature_importance_albation} where statistical and Doppler-based features show 0 or often negative importance. This suggests that the model actively suppresses these features due to their noise sensitivity. In contrast, the earlier selected feature set still retains some level of importance across varying noise conditions. 
Side lobe energy ratio and Temporal Energy Variance represent interesting cases: their importance remains low throughout varying noise levels, yet removing these leads to reduced accuracy and F1 scores. This shows that despite these features are not given significant importance by the model, they provide complementary information aiding the model's classification decision. This points to the fact, that these features still hold discriminative power even if permutation importance remains low.

\begin{table}[!t]
\centering
\caption{Ablation study on feature subsets}
\label{tab:ablation}
\begin{tabular}{p{4.0cm}cccc}
\toprule
Feature Set & Accuracy ($\mu \pm \sigma$) & F1 ($\mu$) & Precision ($\mu$) & Recall ($\mu$) \\
\midrule
Selected feature set & 0.916 $\pm$ 0.060 & 0.912 & 0.947 & 0.892  \\
Full feature set & 0.907 $\pm$ 0.078 & 0.904 & 0.945 & 0.884 \\
With Doppler features & 0.890 $\pm$ 0.080 & 0.889 & 0.917 & 0.874 \\
With statistical features & 0.907 $\pm$ 0.056 & 0.905 & 0.939 & 0.886 \\
Without sidelobe energy ratio & 0.890 $\pm$ 0.065 & 0.889 & 0.915 & 0.875 \\
Without temporal energy variance & 0.898 $\pm$ 0.080 & 0.898 & 0.927 & 0.881 \\
\bottomrule
\end{tabular}
\end{table}

\subsection{Limitations and Future Work}
\begin{itemize}
    \item \textbf{Single dataset evaluation: }Future research can extend the research carried out in this study to include multiple datasets with varying radar and spectrogram parameters. Although the results demonstrate that handcrafted features hold strong discriminative power, the evaluation is limited to a single dataset. Inter-dataset variability and operating conditions for radars may significantly affect the feature distribution.
    
    \item \textbf{Synthetically generated noise}: Simulated noise does not capture the full range of radar signal degradation. Practical radars are affected by varying oscillator jitters, multi path propagation, line of sight obstructions and other hardware induced distortions. These conditions can alter frequency and energy distributions which in turn affect the feature distributions. Hence, the results need to be observed for realistic noise conditions.
    
    \item \textbf{Absence of multiple target per spectrogram:}The  study considers isolated target detection. In practical radar scenarios, the presence of multiple targets introduces challenges, such as micro-Doppler superposition and other interference patterns that might alter feature distributions. Extending the framework to multi-target and clutter-rich environments is necessary to assess operational viability.
    
    \item \textbf{Handcrafted representation limits:} Further research can be carried out to extend the study to include hybrid or learned feature representations. Although handcrafted features capture physically meaningful features, they may not capture higher-order information from the temporal and frequency domains or cross-frequency dependencies. Exploring other models with a trade-off between interpretation and learning capabilities remains an open direction for future work.
\end{itemize}




\section{Conclusion}

This work investigates the robustness of physics-motivated handcrafted features for micro-Doppler based classification of drones, birds, and reflectors under controlled noise conditions. Using a compact feature set, mean classification accuracies of $0.916 \pm 0.095$ for the SVM and $0.916 \pm 0.060$ for the Random Forest classifier were achieved on noise-free data, with macro-averaged F1-scores of $0.909$ and $0.912$, respectively. Beyond baseline performance, the key contribution of this study lies in the robustness analysis of features under AWGN, phase noise, and their combined effects. Rather than exhibiting a sharp drop under extreme noise, the features extracted from the spectrograms degrade gracefully, indicating that they capture structurally stable characteristics of the micro-Doppler spectrogram. Permutation-based analysis revealed that entropy-based descriptors remain relatively dominant across noise regimes and levels. Whereas statistical features such as variance-based and higher-order descriptors exhibit reduced and often unstable importance. This behavior reflects the capability of the underlying physics of the task, entropy-based features capture global dispersion of energy across spectral and temporal marginals that remain relatively stable when the noise floor is raised, while moment-based descriptors depend more strongly on precise amplitude structure.Results from the ablation study reveals that carefully crafted features that encode physically meaningful phenomena can outperform larger feature sets that include redundant or noise-sensitive descriptors. These findings suggest that feature design, guided by robustness analysis and interpretability tools offers a competitive and data-efficient alternative to purely learned representations for radar based micro-Doppler classification, especially under resource-constrained or noisy scenarios. The study emphasizes interpretability and robustness over raw performance maximization, addressing practical deployment considerations in radar systems. These findings are relevant for real-time embedded radar systems where computational and data constraints limit deep learning deployment.


\bibliographystyle{elsarticle-num}
\bibliography{References}

\end{document}